\documentclass[a4paper,10pt,conference]{IEEEtran}

\usepackage{latexsym}
\usepackage{graphicx}
\usepackage[mathscr]{eucal}
\usepackage{cite}
\usepackage{subcaption}
\usepackage{comment}
\usepackage{amsthm}
\usepackage{overpic}
\usepackage{steinmetz}
\usepackage{array}
\usepackage{url}
\usepackage{xcolor}
\IEEEoverridecommandlockouts

\theoremstyle{plain}

\usepackage{latexsym}
\usepackage{graphicx}
\usepackage{amsfonts,amssymb,amsmath}
\usepackage{varwidth}
\usepackage[T1]{fontenc}
\usepackage{cite}
\usepackage{subcaption}
\usepackage{comment}
\usepackage{overpic}
\usepackage{steinmetz}
\usepackage{array}
\usepackage{url}
\usepackage{xcolor}
\usepackage{algorithm,algorithmic}

\IEEEoverridecommandlockouts

\newcommand{\argmax}[1]{{\underset{{#1}}{\mathrm{arg\,max}}}}

\usepackage{fancyhdr}
\begin{document}

\title{\textcolor{black}{Joint Uplink–Downlink Fronthaul Bit Allocation in Fronthaul-Limited Massive MU-MIMO Systems}
\thanks{This work was supported by the Wallenberg Academy Fellow program.}
}

\author{\IEEEauthorblockN{Yasaman Khorsandmanesh, Emil Björnson, and Joakim Jaldén}
\IEEEauthorblockA{
\textit{KTH Royal Institute of Technology, Stockholm, Sweden}\\
Email: \{yasamank, emilbjo, jalden\}@kth.se}}

\maketitle
\begin{abstract}
This paper optimizes the fronthaul bit allocation in massive multi-user multiple-input multiple-output (MU-MIMO) systems operating with limited-capacity fronthaul links. We consider an advanced antenna system (AAS) controlled by a centralized baseband unit (BBU). In the AAS, multiple antenna elements together with their radio units are integrated into a single unit. In this setup, a key challenge is allocating fronthaul bits between uplink channel state information (CSI) quantization and downlink precoding matrix quantization. We formulate the problem of maximizing the sum spectral efficiency (SE) for a given fronthaul capacity. We develop an SE expression for this scenario based on the hardening bound. We compute the expression in closed form for maximum ratio transmission, which reveals the relative impact of the two types of quantization distortion. We then formulate a bit split optimization problem and propose an algorithm that exactly solves it.
 Numerical results demonstrate how the relative importance of assigning bits to CSI and precoding varies depending on the signal-to-noise ratio.
\end{abstract}

\section{Introduction}
The rapid growth in demand for high data rates, reliable connectivity, and large-scale access is a key challenge for 5G wireless networks and beyond. To meet these requirements, massive multiple-input multiple-output (MIMO) has been identified as a key technology, which combines high spectral efficiency with improved energy efficiency, enabled by spatial multiplexing\cite{Marzetta:2010,Larsson:2014}. A typical massive MIMO system has a multi-user MIMO (MU-MIMO) configuration, where a base station (BS) with a large number of antennas communicates with multiple user equipments (UEs) on the same time-frequency resources.

A typical 5G base station (BS) consists of two main components: the advanced antenna system (AAS) and the baseband unit (BBU). The AAS integrates the antenna elements with their corresponding radio units (RUs) into a single physical unit, while the BBU handles the digital processing for both the received uplink data and transmitted downlink data. These two components are interconnected via a digital fronthaul. The integration of antennas and radios into a single box has made massive MU-MIMO systems practically feasible \cite{Bjornson2019d}, and has facilitated the virtualization of the BBU in edge clouds by migrating to a centralized radio access network (C-RAN) architecture \cite{peng2015fronthaul}. However, the main bottleneck in this new implementation arises from the limited fronthaul capacity.
Both uplink/downlink data and the combining/precoding coefficients are transmitted over this fronthaul link and must be quantized to a finite resolution, which leads to performance degradation. In response to this challenge, this paper introduces an algorithm to allocate bidirectional bits optimally to maximize the sum rate under fronthaul capacity limitations.

\subsection{Prior Work}

Much of the work on massive MIMO with limited fronthaul has focused on cell-free networks, where multiple AAS share the same BBU. One approach to addressing the limited fronthaul capacity in these systems is to compress the data exchanged between the central processing unit (CPU) and the access points (APs) via quantization. Several recent works have explicitly considered fronthaul data quantization \cite{masoumi2019performance,kim2022cell}, typically using uniform quantization modeled through the additive quantization noise model (AQNM). For instance, \cite{masoumi2019performance} analyzed the achievable rate under hardware impairments, while \cite{kim2022cell} proposed a codebook that minimizes the channel estimation error in scenarios with both limited fronthaul capacity and low-resolution analog-to-digital/digital-to-analog converters (ADCs/DACs). These studies, however, assume a fixed fronthaul bit allocation for either uplink or downlink scenarios.
Another work that investigated dynamic fronthaul bit allocation is \cite{liu2015joint}. Recently, \cite{rajapaksha2023unsupervised} proposed a machine-learning-based fronthaul capacity allocation scheme, but it focused on quantizing the precoded signals, even though it is more efficient to quantize only the precoding matrix since the data signals have a minimal bit representation
\cite{khorsandmanesh2023optimized}
. \cite{kim2024meta} considers both full-duplex and half-duplex fronthaul links with limited capacity; however, it only models the limited fronthaul as distortion.

Despite these advances, the problem of optimal bit allocation under joint uplink and downlink fronthaul constraints in MU-MIMO remains largely unexplored. In practice, the BBU computes the precoding matrix from quantized channel state information (CSI) received over the uplink and then forwards quantized precoding weights to the AAS in the downlink. Both CSI and precoding coefficients are complex-valued and must be quantized, so efficient bit allocation between these components is essential. This motivates the joint fronthaul bit allocation problem addressed in this work.

\subsection{Contributions}
This paper proposes a joint uplink/downlink quantization bit allocation strategy designed to maximize the sum spectral efficiency (SE), subject to fronthaul capacity constraints. The main contributions of this work are as follows:
\begin{itemize}
  \item We formulate an optimization problem to maximize the overall sum SE, specifically focusing on optimal bit allocation across uplink and downlink transmissions within a fronthaul-limited system. 
  \item For maximum ratio transmission (MRT) precoding, we derive an approximate closed-form signal-to-interference-plus-noise ratio (SINR) expression using the AQNM, emphasizing the distinct effects of CSI and precoding quantization on performance. We also investigate the system behavior with the Zero-Forcing (ZF) and Wiener Filter (WF) precoding schemes. 
  \item We provide numerical results demonstrating the performance of the proposed bit allocation scheme, showing its effectiveness in maximizing SE with a limited fronthaul.
\end{itemize}

\section{System Model}
\begin{figure}[!t]
   \begin{overpic}[width=0.65\columnwidth]{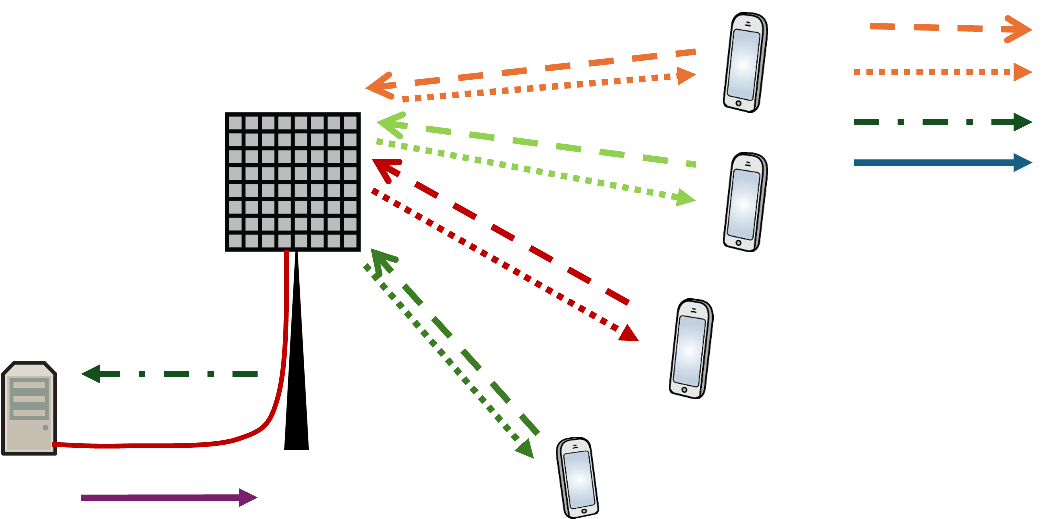}
    \put(-4,-1){\small BBU}%
    \put(5.5,8.5){\scriptsize  Fronthaul}%
    \put(23.5,-1){\small AAS}%
    \put(70,16){\small UEs}%
    \put(99,46){\scriptsize Uplink Pilot Transmission}%
    \put(99,41.5){\scriptsize Downlink Data Transmission}%
    \put(99,37){\scriptsize Channel Estimation}%
    \put(99,33){\scriptsize Precoded Signal}%
   \end{overpic}
   \caption{A massive MU-MIMO system where the AAS estimates uplink channels, forwards quantized CSI to the BBU via a limited-capacity fronthaul, and then the BBU sends the quantized precoding matrix for downlink data transmission towards the AAS.} 
\label{fig:systemmodel}
\end{figure}
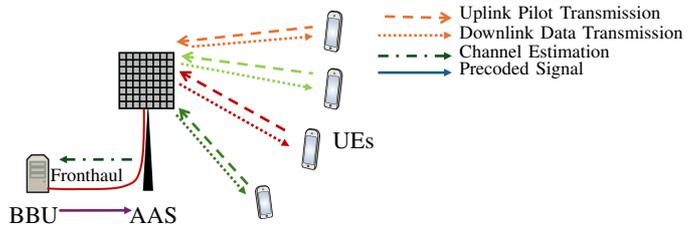

We consider a single-cell massive MU-MIMO system where an AAS equipped with $M$ antennas serves $K \ll M$ single-antenna UEs. The system operates in block-fading time-division duplex (TDD) mode. The uplink channel between the AAS and UE $k$ is modeled as independent Rayleigh fading $\mathbf{h}_{k} \sim \mathcal{CN}(\mathbf{0},\beta_k\mathbf{I}_{M}),$ where $\beta_k$ is the large-scale fading coefficient.  The AAS is connected to the BBU via a fronthaul link of limited capacity $C_{\rm FH}$ [bits/s/Hz], as illustrated in Fig.~\ref{fig:systemmodel}. The AAS estimates the UE channels from uplink pilot transmissions and forwards quantized CSI to the BBU via this link. The BBU then computes a downlink precoding matrix, which is quantized and transmitted to the AAS for downlink data transmission. Since data symbols are already ``quantized'' as channel-coded bits, they can be transmitted over the fronthaul without quantization error, mapped to modulation symbols at the AAS, and then multiplied with the quantized precoding matrix before being transmitted wirelessly to the UEs, which is more efficient in massive MIMO systems \cite{khorsandmanesh2023optimized}. We assume that the AAS can transform the uplink pilot signal into a channel estimate, since it only requires a scalar multiplication.\footnote{For other channel models (e.g., correlated Rayleigh fading) where the estimator is more complex, it might be necessary to implement the estimation at the BBU. This case is deferred to future research.}

\subsection{Uplink Channel Estimation and Quantization}
In TDD operation, the AAS estimates the uplink channels and sends them to the BBU, which uses them for both uplink combining and downlink precoding, by exploiting channel reciprocity~\cite{demir2021foundations}. In each coherence block of length $\tau_{\mathrm{c}}$, UE $k$ transmits a pilot sequence $\boldsymbol{\phi}_{k} \in \mathbb{C}^{\tau_{\mathrm{\rm P}}}$ of length $\tau_{\mathrm{\rm P}}$, leading to a pilot overhead of $\tau_{\mathrm{\rm P}}/\tau_{\mathrm{c}}$. The pilot sequences are orthonormal, i.e., $\boldsymbol{\phi}_{i}^{\rm H}\boldsymbol{\phi}_{j} = 0$ for $i\neq j$ and $\boldsymbol{\phi}_{i}^{\rm H}\boldsymbol{\phi}_{i} = 1$, which requires $\tau_{\mathrm{\rm P}} \geq K$. Each UE transmits $\sqrt{\tau_{\mathrm{\rm P}}}\,\boldsymbol{\phi}_{k}^{\rm H}$ to guarantee that the total pilot energy scales proportionally with the pilot length.  

The received uplink pilot signal at the AAS is  
\begin{equation}  
    \mathbf{Y}^{\text{pilot,U}} = \sum_{k=1}^{K}\sqrt{q_{k}\tau_{\mathrm{\rm P}}}\,\mathbf{h}_{k}\boldsymbol{\phi}_{k}^{\rm H} + \mathbf{N}^{\text{pilot,U}},
\end{equation}
where $q_{k}$ is the transmit power used by the UE, normalized by the noise power, and $\mathbf{N}^{\text{pilot,U}} \in \mathbb{C}^{M \times \tau_{\mathrm{\rm P}}}$ has i.i.d. $\mathcal{CN}(0,1)$ entries. After multiplying the received signal $\mathbf{Y}^{\text{pilot,U}}$ with the pilot vector $\boldsymbol{\phi}_{k}$ of $k$-th UE from the left, we obtain  
\begin{equation}   
    \mathbf{y}_{k}^{\rm U} = \sqrt{q_{k} \tau_{\mathrm{\rm P}}}\,\mathbf{h}_{k} + \mathbf{n}_{k}^{\rm U},
\end{equation}
with $\mathbf{n}_{k}^{\rm U} = \mathbf{N}^{\text{pilot,U}}\boldsymbol{\phi}_{k} \sim \mathcal{CN}(\mathbf{0},\mathbf{I}_{M})$. The MMSE estimate of the uplink channel vector $\mathbf{h}_{k}$ is \cite[Ch.~3]{bjornson2017}  
\begin{equation}  
    \hat{\mathbf{h}}_{k} = \frac{\beta_k}{{q}_{k}\tau_{\mathrm{\rm P}} \beta_k + 1}\mathbf{y}_{k}^{\rm U},
    \label{eq:mmseestimate}
\end{equation}
with the covariance matrix
\begin{equation}
\label{eq:Chat}
    \mathbf{C}_{\hat{h}_k}
    = \gamma_k \mathbf{I}_{M},
\end{equation}
 where $\gamma_k = \frac{q_k \tau_{\mathrm{P}} \beta_k^2}{q_k \tau_{\mathrm{P}} \beta_k + 1}$.
 
 The AAS then transmits the estimated uplink channel matrix $\hat{\mathbf{H}}=[\hat{\mathbf{h}}_{1},\dots,\hat{\mathbf{h}}_{K}] \in \mathbb{C}^{M \times K}$ to the BBU, which leads to quantization errors over the fronthaul. We utilize the AQNM \cite{orhan2015low} approximation by adding an additive white noise that models the quantization distortion of each channel entry. Hence, the quantized CSI that reaches the BBU is modeled as
\begin{equation}
  \hat{\mathbf{H}}_{\rm Q} = (1 - \eta_{\rm H})\hat{\mathbf{H}} + \mathbf{N}_{\mathrm{Q},H},
    \label{eq:CQCSI}
\end{equation}
where $\eta_{\rm H}$ is the AQNM distortion factor, which is a monotonically decreasing function of $B_{\rm H}$ (bits per complex entry). For the Gaussian input distribution that we get under Rayleigh fading, the value of $\eta_{\rm H}$ for $B_{\rm H} \le 5$ is listed in Table~\ref{tab:eta_h} at the top of the next page and for  $B_{\rm H} > 5$ it can be tightly approximated by $\eta_{\rm H} = \frac{\pi\sqrt{3}}{2} \cdot 2^{-2B_{\rm H}}$ \cite{fan2015uplink}. The quantization noise  $\mathbf{N}_{\mathrm{Q},H}$ in \eqref{eq:CQCSI} is uncorrelated, has zero mean, and covariance $ \eta_{\rm H}(1-\eta_{\rm H})\mathbb{E}[|\hat{h}_{m,k}|^2])$ per entry, for $m=1,\ldots,M$ and $k=1,\ldots,K$. While the quantization noise is, in practice, slightly correlated across antennas and UEs, prior results show that if the number of quantization bits is practically large, modeling it as independent and identically distributed (i.i.d.) yields accurate performance predictions~\cite{bin2017bit}. We will numerically verify the accuracy of this approximation in Section~\ref{sec:results}. The covariance matrix of the quantized CSI at the BBU is
\begin{equation}
   \mathbf{C}_{\hat{h}_{{\rm Q}_k}}=(1-\eta_{\rm H})^2\mathbf{C}_{\hat{h}_k}+\mathbf{C}_{{N}_{\mathrm{Q},\hat{h}_{k}}},
\end{equation}
where $\mathbf{C}_{{N}_{\mathrm{Q},\hat{h}_{k}}}$ is a diagonal matrix that contains the diagonal of $\mathbf{C}_{\hat{h}_k}$ in \eqref{eq:Chat}. 

\begin{table}[t!]
\centering
\caption{AQNM distortion $\eta_{\rm H}$  vs. $B_{\rm H}$.}
\label{tab:eta_h}
\begin{tabular}{|c|ccccc|}
\hline
$B_{\rm H}$      & $1$      & $2$       & $3$       & $4$        & $5$      \\
\hline
$\eta_{\rm H}$  & $0.3634$ & $0.1175$  & $0.03454 $& $0.009497$ & 0.002499  \\
\hline
\end{tabular}
\end{table}
\subsection{Precoder Design and Downlink Quantization}

The BBU computes a linear precoding matrix $\mathbf{ P}\in\mathbb{C}^{M\times K}$ based on the downlink channel $\hat{\mathbf{H}}_{\rm Q, D} \in \mathbb{C}^{K \times M}$, which is $\hat{\mathbf{H}}_{\rm Q, D} = \hat{\mathbf{H}}_{\rm Q}^{\rm T}$ due to the TDD and channel reciprocity assumption. We consider three standard linear precoders:

\subsubsection*{1) Maximum Ratio Transmission (MRT)} This scheme aligns each precoder with the estimated channel to maximize the received signal-to-noise ratio(SNR), deliberately ignoring inter-user interference, which yields 
\begin{equation}  
\label{eq:MRT}
\mathbf{ P}_{\mathrm{MRT}} = \zeta_{\mathrm{MRT}} \, \hat{\mathbf{H}}_{\rm Q,D}^{\rm H},
\end{equation}
where $\zeta_{\mathrm{MRT}} = \sqrt{\frac{P_{\rm t}}{\text{tr}(\hat{\mathbf{H}}_{\rm Q,D}\hat{\mathbf{H}}_{\rm Q,D}^{\rm H})}}$ is the precoding factor and $P_{\rm t}$ is the average downlink transmit power.

\subsubsection*{2) Zero-Forcing (ZF)} This scheme chooses precoders that null inter-user interference and is obtained by 
\begin{equation}
    \mathbf{ P}_{\mathrm{ZF}} = \zeta_{\mathrm{ZF}} \hat{\mathbf{H}}_{\rm Q,D}^{\rm H}\left(\hat{\mathbf{H}}_{\rm Q,D}\hat{\mathbf{H}}_{\rm Q,D}^{\rm H}\right)^{-1},
\end{equation}
where $\zeta_{\mathrm{ZF}} = \sqrt{\frac{P_{\rm t}}{\text{tr}\Big( (\hat{\mathbf{H}}_{\rm Q,D}\hat{\mathbf{H}}_{\rm Q,D}^{\rm H})^{-1}\Big) }}$.

\subsubsection*{3) Wiener Filter (WF)} This precoder is a kind of regularized ZF (RZF) that minimizes the mean square error under a sum-power constraint, trading exact interference nulling against noise amplification. The precoding matrix is calculated as 
\begin{equation}
    \mathbf{ P}_{\mathrm{WF}} = \zeta_{\mathrm{WF}}  \hat{\mathbf{H}}_{\rm Q,D}^{\rm H}\left(\hat{\mathbf{H}}_{\rm Q,D}\hat{\mathbf{H}}_{\rm Q,D}^{\rm H} + \frac{K\sigma^2}{P_{\rm t}} \mathbf{I}_{K}\right)^{-1},
\end{equation}
where $\sigma^2$ is the downlink noise variance and 
\begin{align}
  \zeta_{\mathrm{WF}} & = \sqrt{{P_{\rm t}}} / \nonumber  \text{tr} \Big((\hat{\mathbf{H}}_{\rm Q,D}\hat{\mathbf{H}}_{\rm Q,D}^{\rm H} + \frac{KN0}{P_{\rm t}} \mathbf{I}_{K})^{-1} \\ &  \hat{\mathbf{H}}_{\rm Q,D}\hat{\mathbf{H}}_{\rm Q,D}^{\rm H}(\hat{\mathbf{H}}_{\rm Q,D}\hat{\mathbf{H}}_{\rm Q,D}^{\rm H} + \frac{K\sigma^2}{P_{\rm t}} \mathbf{I}_{K})^{-1} \Big)^{1/2} . \nonumber
\end{align}

When the BBU has computed the precoding matrix $\mathbf{P}$, it sends it over the limited-capacity fronthaul towards the AAS, which leads to further quantization errors. By applying the AQNM, we can write the quantized precoder at the AAS as
\begin{equation}
    \mathbf{ P}_{\rm Q} = (1 - \eta_{\rm P})\mathbf{ P} + \mathbf{N}_{\rm Q , P},
    \label{eq:quantisedprecod}
\end{equation}
where $\eta_{\rm P}$ is calculated as $\eta_{\rm H}$ with $B_{\rm P}$ bits per complex entry, and $\mathbf{N}_{\mathrm{Q},P}$ is the corresponding uncorrelated quantization noise with zero mean and per entry variance $\eta_{\rm P}(1-\eta_{\rm P})\mathbb{E}[|{\rm p}_{k,m}|^2]$ for $m=1,\ldots,M$ and $k=1,\ldots,K$, where ${\rm p}_{k,m}$ is the $(k,m)$-th entry of $\mathbf{P}$. Hence, the $k$-th column of the noise matrix is distributed as $\mathbf{n}_{\mathrm{Q},p_k} \sim \mathcal{CN}(\mathbf{0}_M,\mathbf{C}_{{N}_{\mathrm{Q},P_{k}}})$, where $ \mathbf{C}_{{N}_{\mathrm{Q},p_{k}}} = \eta_{\rm P} (1 - \eta_{\rm P}) \mathbf{D}_{p_k}$
and
\begin{equation}
    \mathbf{D}_{{p}_k} \triangleq \operatorname{diag}\!\big(\mathbb{E}\{|p_{k,1}|^2\},\ldots,\mathbb{E}\{|p_{k,M}|^2\}\big).
\end{equation}

\subsection{Downlink Transmission}

The downlink data symbols correspond to coded bit sequences from a codebook and can be transferred over the fronthaul without introducing quantization errors. At the AAS, the UEs’ data symbols are mapped to modulation symbols, and the quantized precoding matrix is then applied to these data symbols. The resulting signals are then transmitted over
the wireless channel towards the UEs using the transmit power $P_{\rm t}$.

Let $\mathbf{s} = [s_1, \ldots, s_K]^\mathrm{T}\in \mathcal{O}^{K}$ ($\mathcal{O}$  ($\mathcal{O}$ is a finite set of constellation points such as a QAM alphabet), denote the vector of downlink data symbols with $\mathbb{E}[\mathbf{s}\mathbf{s}^{H}] = \mathbf{I}_{K}$. The AAS transmits the precoded signal
\begin{equation}
    \mathbf{x} = \alpha \mathbf{ P}_{\rm Q}\mathbf{s},
\end{equation}
    where $\alpha = \sqrt{\frac{P_{\rm t}}{\lVert \mathbf{ P}_{\rm Q} \rVert_{\mathrm{F}}^{2}}}$ is calculated at the AAS to ensure full transmit power, since $\lVert \mathbf{ P} \rVert_{\mathrm{F}}^{2}= P_{\rm t}$  does not, in general, imply
$\lVert \mathbf{ P}_{\rm Q} \rVert_{\mathrm{F}}^{2}= P_{\rm t}$.
The received signal at the UEs is  
\begin{equation}
    \mathbf{y} = \mathbf{H}_{\rm D}\mathbf{x} + \mathbf{n} 
    = \alpha\mathbf{H}_{\rm D}\mathbf{ P}_{\rm Q}\mathbf{s} + \mathbf{n}, \label{eq:recivedDL}
\end{equation}
where $\mathbf{y} = [y_1,\ldots,y_K]^\mathrm{T}$,  $\mathbf{H}_{\rm D} = [\mathbf{h}_1^\mathrm{T},\ldots,\boldsymbol{h}_K^\mathrm{T}]^\mathrm{T}$ is the downlink channel matrix, the precoding matrix can be expressed as $\mathbf{ P}_{\rm Q} = [\mathbf{p}_{{\rm Q}_1},\ldots,\mathbf{p}_{{\rm Q}_K}]$, and $\mathbf{n}\sim\mathcal{CN}(\mathbf{0},\sigma^2\mathbf{I}_{K})$ is an additive white Gaussian noise vector.

\subsection{Achievable Downlink SE: Hardening Bound}

To quantify the communication performance, we will now derive an achievable downlink SE expression for the signal model in \eqref{eq:recivedDL}. We use the hardening bound that is widely used in the Massive MIMO literature to compute SEs \cite{bjornson2017}. To this end, the received signal at UE $k$ is expanded as  
\begin{align}
y_{k}
&= \underbrace{  \mathbb{E}\{ \alpha \mathbf{h}_{k}^{\rm T} \mathbf{ p}_{{\rm Q}_k} \} s_{k} }_{\text{Desired signal over average channel}}  + \underbrace{\alpha \big( \mathbf{h}_{k}^{\rm T} \mathbf{p}_{{\rm Q}_k} - \mathbb{E}\{ \alpha\mathbf{h}_{k}^{\rm T} \mathbf{ p}_{{\rm Q}_k} \} \big) s_{k} }_{\text{Random channel variation}}   \nonumber \\
&\quad  + \underbrace{ \alpha \sum_{\substack{i=1 \\ i \neq k}}^{K} \mathbf{h}_{k}^{\rm T} \mathbf{p}_{{\rm Q}_i} s_{i} }_{\text{Multi-user interference}} + \underbrace{ n_{k} }_{\text{Noise}}. 
\label{eq:recivedsignalUE}
\end{align} 
The first term in \eqref{eq:recivedsignalUE} represents the desired signal transmitted over the deterministic average precoded channel $\mathbb{E}\{ \mathbf{h}_{k}^{\rm T} \mathbf{ p}_{{\rm Q}_k} \}$, whereas the remaining terms correspond to variables with realizations that are unknown to the UE. By treating the random term $\alpha \big( \mathbf{h}_{k}^{\rm T} \mathbf{ P}_{{\rm Q}_k} - \mathbb{E}\{ \mathbf{h}_{k}^{\rm T} \mathbf{ P}_{{\rm Q}_k} \} \big) s_{k} $ and the interference in \eqref{eq:recivedsignalUE} as uncorrelated noise, an achievable SE for UE $k$ is
\begin{equation}
    \text{SE}_k = \left(1 - \frac{\tau_{\mathrm{\rm P}}}{\tau_{\mathrm{c}}}\right)
    \log_2\left(1 + \Gamma_k\right) \quad [\text{bit/s/Hz}],
    \label{eq:se}
\end{equation}
where the effective SINR is given by
\begin{equation}
    \Gamma_k = 
    \frac{\left|\mathbb{E}\{ \alpha \mathbf{h}_k^{\rm T}\mathbf{ P}_{{\rm Q}_k}\}\right|^2}{ \sum_{i=1}^{K}  \, \mathbb{E}\{ | \alpha  \mathbf{h}_{k}^{\rm T} \mathbf{ P}_{{\rm Q}_i}|^2 \}
-  \big| \mathbb{E}\{  \alpha  \mathbf{h}_{k}^{\rm T} \mathbf{ P}_{{\rm Q}_k} \} \big|^2
+ \sigma^2   }.
\label{eq:hardening-SINR}
\end{equation}

\section{Problem Formulation and Solution Approach} 

We assume the fronthaul link is half-duplex, allowing the fronthaul capacity to be dynamically shifted between uplink and downlink.
We want to find the optimal bit allocation $B_{\rm H}$ and $B_{\rm P}$ that maximizes the sum SE subject to the fronthaul-capacity constraints.
This optimization problem is formulated as
\begin{align}
    \underset{\,B_{\rm H},\,B_{\rm P}}{\text{maximize}} \quad 
    & \sum_{k=1}^K 
    \text{SE}_k(B_{\rm H},\,B_{\rm P}) \label{eq:opt-prob}\\
    \text{subject to}\quad 
    & (B_{\rm H}+B_{\rm P})KM +\nonumber \\ &   ( B_{\rm s}^{\rm UL} T_{\rm u}  B_{\rm s}^{\rm DL} T_{\rm d} )K\;\le\; C_{\mathrm{FH}}, \label{eq:fronthaul-constraint}\\
    & B_{\rm H} \ge 1,\;\; B_{\rm P} \ge 1,\;\; 
\end{align}
where $\text{SE}_k$ is given by \eqref{eq:se} and now treated as a function of $B_{\rm H},\,B_{\rm P}$. The fronthaul constraint \eqref{eq:fronthaul-constraint} accounts for CSI exchange and precoding signaling (both scaling with $MK$), the transmission of uplink detected symbols at a rate $B_{\rm s}^{\rm UL}$ bits per symbol over $T_{\rm u}$ uplink payload symbols per coherence block and the delivery of UE data symbols at a rate $B_{\rm s}^{\rm DL}$ bits per symbol over $T_{\rm d}$ downlink payload symbols per coherence block. This formulation explicitly captures the joint effect of pilot overhead, fronthaul quantization of CSI and precoding, and downlink transmission on the achievable SE with a half-duplex fronthaul link.

\begin{algorithm}[t]
\caption{Exact bit split via finite line search}
\label{alg:linesearch}
\begin{algorithmic}[1]
\STATE \textbf{Input:} $\{{\text{SE}}_k\}$, $P_{\rm t}$, $C_{\rm FH}$, $K$,$M$,$T_{\rm d}$,$B_{\rm s}$; choose a precoding scheme (e.g., MRT/ZF/WF).
\STATE Compute $\bar B$ from \eqref{eq:Bbar}; set best value $v^\star\!\leftarrow\!-\infty$.
\FOR{$B_{\rm H}=1,\ldots,\bar B-1$}
  \STATE $B_{\rm P}\leftarrow \bar B-B_{\rm H}$; form $\eta_{\rm H}$, $\eta_{\rm P}$ in Table~\ref{tab:eta_h}.
  \STATE Evaluate $\Gamma_k$ via \eqref{eq:hardening-SINR} ; compute $\text{SE}_k$ using \eqref{eq:se}.
  \STATE $v\leftarrow \sum_k  \text{SE}_k$; \textbf{if} $v>v^\star$ \textbf{then} update $(B_{\rm H}^\star,B_{\rm P}^\star)\!\leftarrow\!(B_{\rm H},B_{\rm P})$, $v^\star\!\leftarrow v$.
\ENDFOR
\STATE \textbf{Output:} $(B_{\rm H}^\star,B_{\rm P}^\star)$ and the corresponding SEs.
\end{algorithmic}
\end{algorithm}

\subsection{Exact bit split by line search (finite search)}\label{sec:solution}
The fronthaul constraint \eqref{eq:fronthaul-constraint} implies a per-entry budget of
\begin{align}
  \label{eq:Bbar}
  \bar B & \triangleq \left\lfloor \frac{C_{\rm FH}-( B_{\rm s}^{\rm UL} T_{\rm u}  + B_{\rm s}^{\rm DL} T_{\rm d} )K}{KM} \right\rfloor, \\ \nonumber & \qquad
   B_{\rm H}+B_{\rm P}\le \bar B,\;\; B_{\rm H},B_{\rm P}\in\mathbb{N},
\end{align}
where $\lfloor x \rfloor$ returns the greatest integer less than or equal to $x$, and $\mathbb{N}$ is the set of natural numbers. As the feasible set is \emph{finite and one-dimensional}, it suffices to consider $B_{\rm H}\in\{1,\ldots,\bar B-1\}$ and then set $B_{\rm P}=\bar B-B_{\rm H}$.

Since only a single integer $B_{\rm H}$ must be selected, the problem in \eqref{eq:opt-prob} can be solved \emph{exactly} by an exhaustive search:
\begin{equation}
  (B_{\rm H}^{\star},B_{\rm P}^{\star}) 
  \in \argmax{B_{\rm H}\in\{1,\ldots,\bar B-1\}} 
  \sum_{k=1}^K \, \text{SE}_k\big(B_{\rm H},\,\bar B-B_{\rm H}\big).
  \label{eq:line-search}
\end{equation}
Algorithm \ref{alg:linesearch} presents this approach.

\subsubsection*{1) Maximum Ratio Transmission (MRT)}

We now specialize the hardening-bound SINR in~\eqref{eq:hardening-SINR} to the case where the \emph{downlink precoding} is MRT, as defined in \eqref{eq:MRT}. We will find a closed-form approximation by treating the quantization noise as an independent additive term. The tightness of this approximation is demonstrated in Section~\ref{sec:results}. Based on the AQNM model in \eqref{eq:CQCSI} and \eqref{eq:quantisedprecod}, we can rewrite \eqref{eq:recivedDL} as 
\begin{align}
   & \mathbf{y}^{\text{MRT}}  = \alpha\mathbf{H}_{\rm D}\mathbf{ P}_{\rm Q}\mathbf{s} + \mathbf{n} = \alpha\mathbf{H}_{\rm D}\Big((1 - \eta_{\rm P})\mathbf{ P} + \mathbf{N}_{\rm Q , P} \Big) \mathbf{s} + \mathbf{n} \nonumber \\ & = \alpha\mathbf{H}_{\rm D}\Big((1 - \eta_{\rm P})\zeta_{\mathrm{MRT}} \, \hat{\mathbf{H}}_{\rm Q,D}^{\rm H}  + \mathbf{N}_{\rm Q , P} \Big) \mathbf{s} + \mathbf{n}
    \nonumber \\ & = \alpha\mathbf{H}_{\rm D}\Bigg((1 - \eta_{\rm P})\zeta_{\mathrm{MRT}} \Big( (1 - \eta_{\rm H})\hat{\mathbf{H}}^{\rm H}_{\rm D} + \mathbf{N}_{\mathrm{Q},H}^* \Big) + \mathbf{N}_{\rm Q , P} \Bigg) \mathbf{s} \nonumber \\ & \quad + \mathbf{n}
    \nonumber \\ & = \alpha \zeta_{\mathrm{MRT}} (1 - \eta_{\rm P}) (1 - \eta_{\rm H}) \mathbf{H}_{\rm D} \hat{\mathbf{H}}^{\rm H}_{\rm D}\mathbf{s} \nonumber \\ & \quad + \alpha \zeta_{\mathrm{MRT}} (1 - \eta_{\rm P}) \mathbf{H}_{\rm D} \mathbf{N}_{\mathrm{Q},H}^* \mathbf{s} + \alpha \mathbf{H}_{\rm D} \mathbf{N}_{\mathrm{Q},P} \mathbf{s} + \mathbf{n}.
    \label{eq:recivedDLMRT}
\end{align}
By utilizing \eqref{eq:hardening-SINR} and under the assumption of independent noise, the hardening bound turns into \eqref{eq:qauntisedsinr_mrt} at the top of the next page. {}Here, we made a large system assumption ($M \gg K$), so 
\begin{equation}
\bar{\zeta}_{\mathrm{MRT}} = \sqrt{\frac{P_{\rm t}}{\text{tr}\Big(\mathbb{E}\{\hat{\mathbf{H}}_{\rm Q,D}\hat{\mathbf{H}}_{\rm Q,D}^{\rm H}\}\Big)}}    
\end{equation}
and
\begin{equation}
    \bar\alpha = \sqrt{\frac{P_{\rm t}}{\mathbb{E}\{ \lVert  \mathbf{ P}_{\rm Q}) \rVert_{\mathrm{F}}^{2}\} }}.
\end{equation}

The closed-form expression in  \eqref{eq:qauntisedsinr_mrt} can be used to achieve a computationally efficient implementation of the bit-split optimization problem in~\eqref{eq:line-search}. Each function evaluation only involves traces of covariance matrices, such as $\operatorname{tr}(\mathbf{C}_{\hat h_k})$ and $\operatorname{tr}(\mathbf{C}_{\hat h_{Q_k}})$. 
With this closed-form formula, no matrix inversion or per-realization precoding update is required. 
In contrast, if WF or ZF precoding is used, the expectations in the SINR expression must be evaluated using Monte--Carlo averaging for every candidate bit pair $(B_{\rm H},B_{\rm P})$, and each trial involves forming and inverting a $K\times K$ Gram matrix. Consequently, the proposed approximation enables a substantially faster search over the discrete bit-allocation pairs. The simulations reported in the next section indicate that the optimal bit resolution that maximizes the sum SE remains practically identical when replacing MRT with WF or ZF precoding.

\begin{figure*}
\begin{align}
\label{eq:qauntisedsinr_mrt}
& \Gamma_k^{\rm MRT} \approx \nonumber
\\ &
\frac{\bar\alpha^2\bar{\zeta}_{\rm MRT}^2 (1-\eta_{\rm P})^2(1-\eta_{\rm H})^2\big|\text{tr}(\mathbf{C}_{\hat{h}_k})\big|^2}{
\begin{aligned}
&\bar\alpha^2\bar{\zeta}_{\rm MRT}^2(1-\eta_{\rm P})^2\Bigg( (1-\eta_{\rm H})^2\Big(\operatorname{tr}(\mathbf{C}_{\hat{h}_k}\mathbf{C}^{\rm H}_{\hat{h}_k}) - \big|\text{tr}(\mathbf{C}_{\hat{h}_k})\big|^2 \Big) + \sum_{\substack{i=1 }}^{K} \operatorname{tr}(\mathbf{C}_{{h}_k}\mathbf{C}_{\hat{h}_{{\rm Q}_i}} ) \Bigg )  +
\bar\alpha^2\eta_{\rm P}(1-\eta_{\rm P}) \beta_k \sum_{i=1}^{K} \operatorname{tr}( \mathbf{D}_{{p}_i})
+\sigma^2 
\end{aligned}}
\end{align}
\hrule
\end{figure*}

The closed-form expression in \eqref{eq:qauntisedsinr_mrt} also clarifies how the dominant impairment changes with the SNR. 
At low SNR, the denominator is dominated by the thermal-noise term, and the system operates in a noise-limited regime. 
In this case, improving the CSI accuracy---that is, reducing $\eta_{\rm H}$ by allocating more bits to $B_{\rm H}$---provides the largest performance gain, while the effect of precoder quantization remains minor. 
As the SNR increases, the quantization-related terms become dominant and the system gradually transitions into a quantization-limited regime. 
Here, additional transmit power yields little improvement, and performance depends primarily on the quantization resolutions. 
Optimal operation is therefore achieved by jointly reducing $\eta_{\rm H}$ and $\eta_{\rm P}$ (increasing $B_{\rm H}$ and $B_{\rm P}$) under the total bit budget$~\bar B$. We will confirm these observations numerically in the next section.

\begin{figure}[!t]
    \centering
\includegraphics[width=\linewidth]{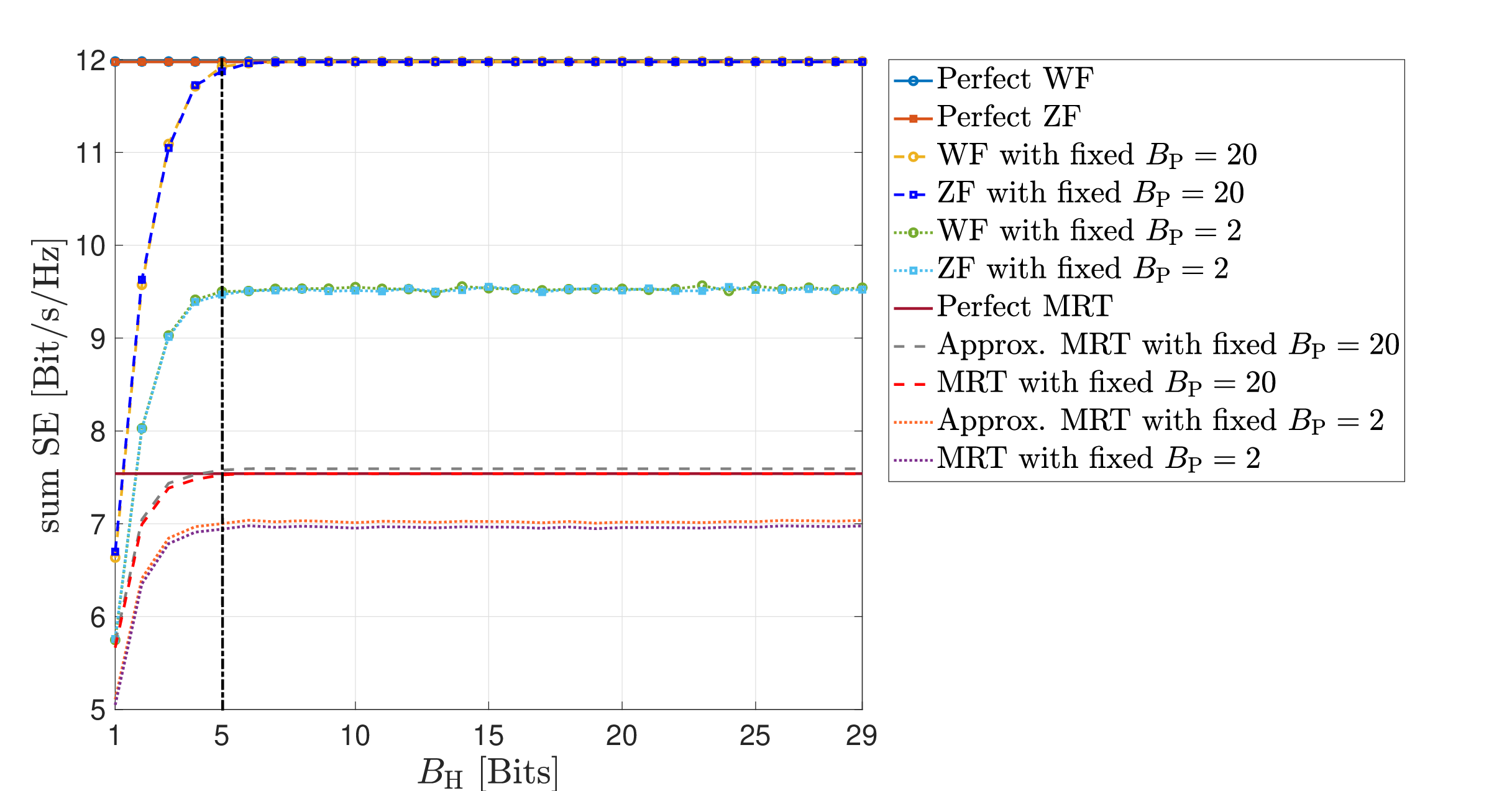}
    \caption{Sum SE versus $B_{\rm H}$ for fixed $B_{\rm P}$ at $\text{SNR}=10$ dB.}
    \label{fig:split}

\end{figure}

\section{Numerical Results}\label{sec:results}

In this section, we will evaluate the sum SE of the system with different precoding schemes under varying conditions.

We present the sum SEs that are evaluated by Monte Carlo simulation with  $1000$ independent trials. Unless otherwise stated, we consider $M=128$ antennas, $K=8$ UEs, coherence block length $\tau_c=200$, and pilot length $\tau_p=K$. The UEs have a common SNR that we define as $\text{SNR}=\frac{P_{\rm t}}{\sigma^2}$. In some simulations, we will deviate from the baseline scenario by considering varying SNRs.

Fig.~\ref{fig:split} depicts the sum SE as a function of $B_{\rm H}$ at $\text{SNR}=10$ dB for the following schemes:
\begin{enumerate}
\item Perfect WF, ZF, and MRT, which considers perfect CSI and corresponding precoding without quantization;
\item WF, ZF, and MRT, with fixed $B_{\rm P}$, which $B_{\rm H}$ varies, but the resolution for precoding is fixed;
\item Approx. MRT, where SE expression for MRT in \eqref{eq:qauntisedsinr_mrt} is used.
\end{enumerate}
The budget is $\bar{B} = 30$ and $B_{\rm H}\in\{1,\ldots,\bar B-1\}$.  As expected, at this SNR, WF and ZF outperform MRT because they actively suppress inter-user interference. The perfect WF/ZF curves form an upper bound around $12$~bit/s/Hz. With $B_{\rm P}=20$ bits, both WF and ZF nearly reach this bound once $B_{\rm H}\!\ge\!5$. Reducing the precoding resolution to $B_{\rm P}=2$ creates a clear gap to the upper bound, essentially independent of $B_{\rm H}$ after the threshold, since coarse precoding leaves residual multi-user interference. MRT is interference-limited even with perfect CSI; correspondingly, its SE saturates around $7.5$~bit/s/Hz. The analytic \emph{Approx.\ MRT} based on \eqref{eq:qauntisedsinr_mrt} closely tracks the simulated MRT for both $B_{\rm P}=20$ and $B_{\rm P}=2$, confirming the tightness of the closed-form SE approximation and its utility for bit-split design. Therefore, at moderate/high SNR, a robust rule-of-thumb for fixed $B_{\rm P}$ is $B_{\rm H}^\star \approx 5\ \text{bits} $. This reflects that CSI quantization is the primary bottleneck only up to a small threshold; beyond that, the precoding resolution dominates the gap to the upper bound. The same conclusions hold when we instead fix $B_{\rm H}$ and vary $B_{\rm P}\in\{1,\ldots,\bar B-1\}$. We also repeated the simulations under spatially correlated Rayleigh channels and saw similar trends, which validates the applicability of the proposed method for more comprehensive setups. A full characterization of the optimal bit split as a function of the eigenvalue spread and inter-user covariance overlap is left for future work. For the rest of the MRT curves, we utilize \eqref{eq:qauntisedsinr_mrt}.

\begin{figure}[!t]
    \centering
\includegraphics[width=0.8\linewidth]{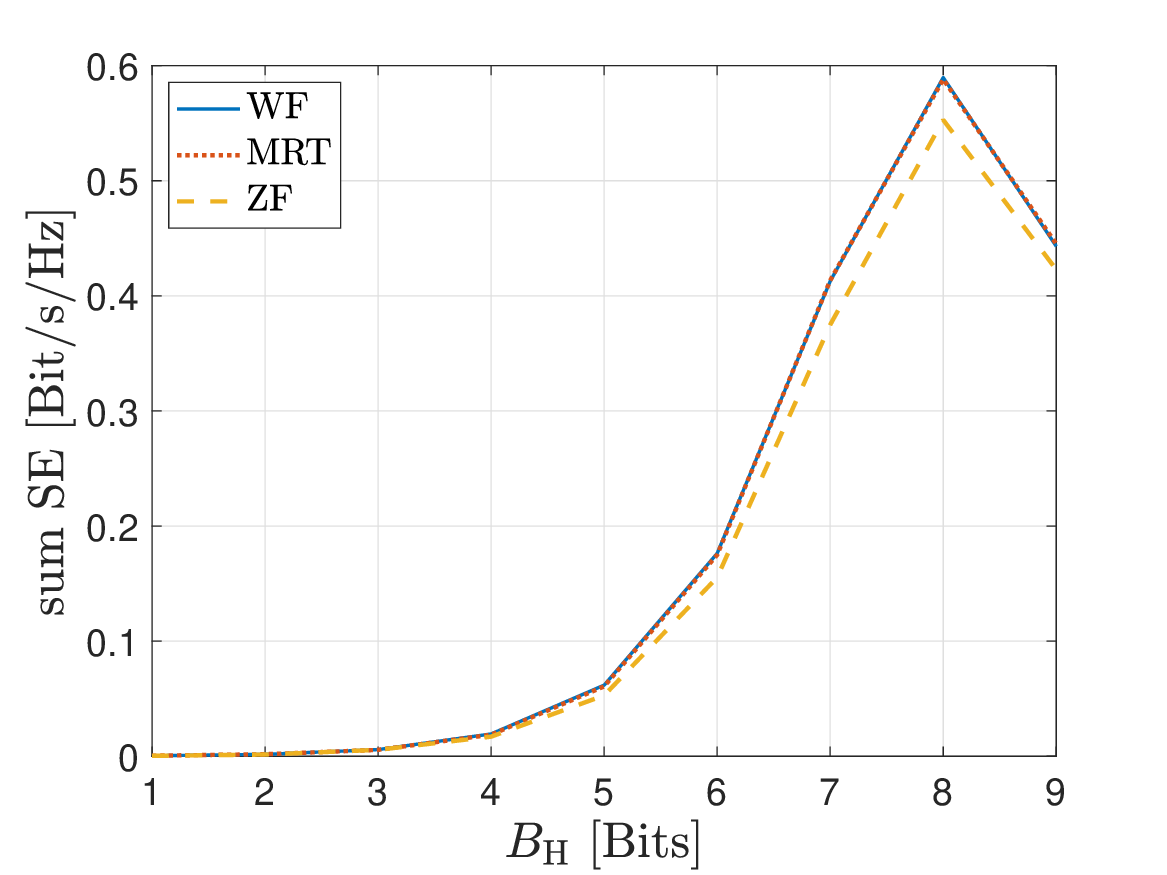}
    \caption{Sum SE versus $B_{\rm H}$ at $\text{SNR}=-15$ dB.}
    \label{fig:lowsnr}
\end{figure}

Fig.~\ref{fig:lowsnr} presents the downlink sum SE versus the CSI–bit allocation $B_{\rm H}$ for $\text{SNR}=-15$\,dB and the total fronthaul budget $\bar B=10$ (so $B_{\rm P}=\bar B-B_{\rm H}$). All curves show a distinct single maximum at $B_{\rm H}^{\star}\!\approx\!8 \quad ( B_{\rm P}\!\approx\!2),$ that is, allocating roughly $80\%$ of the budget to transfer CSI to the BBU and the remainder to transfer the computed precoder back to the AAS. For $B_{\rm H}<B_{\rm H}^{\star}$, the left-hand slope is due to  CSI inaccuracy, where too few CSI bits cause significant channel mismatches, so both the coherent array gain and interference mitigation are severely impaired. For $B_{\rm H}>B_{\rm H}^{\star}$ the SE drops because $B_{\rm P}$ becomes too small; the coarse precoding generates quantization distortion and residual multi-user interference, which dominate in spite of improved CSI (the effect is particularly happened at $B_{\rm H}=9$, i.e., $B_{\rm P}=1$). WF and MRT nearly coincide and outperform ZF, which is expected at low SNR. The small gap between WF and MRT reflects that, in the noise-limited regime with tight fronthaul, interference suppression provides limited benefit once the CSI is sufficiently coarse quantized.

\begin{figure}[!t]
    \centering
\includegraphics[width=0.8\linewidth]{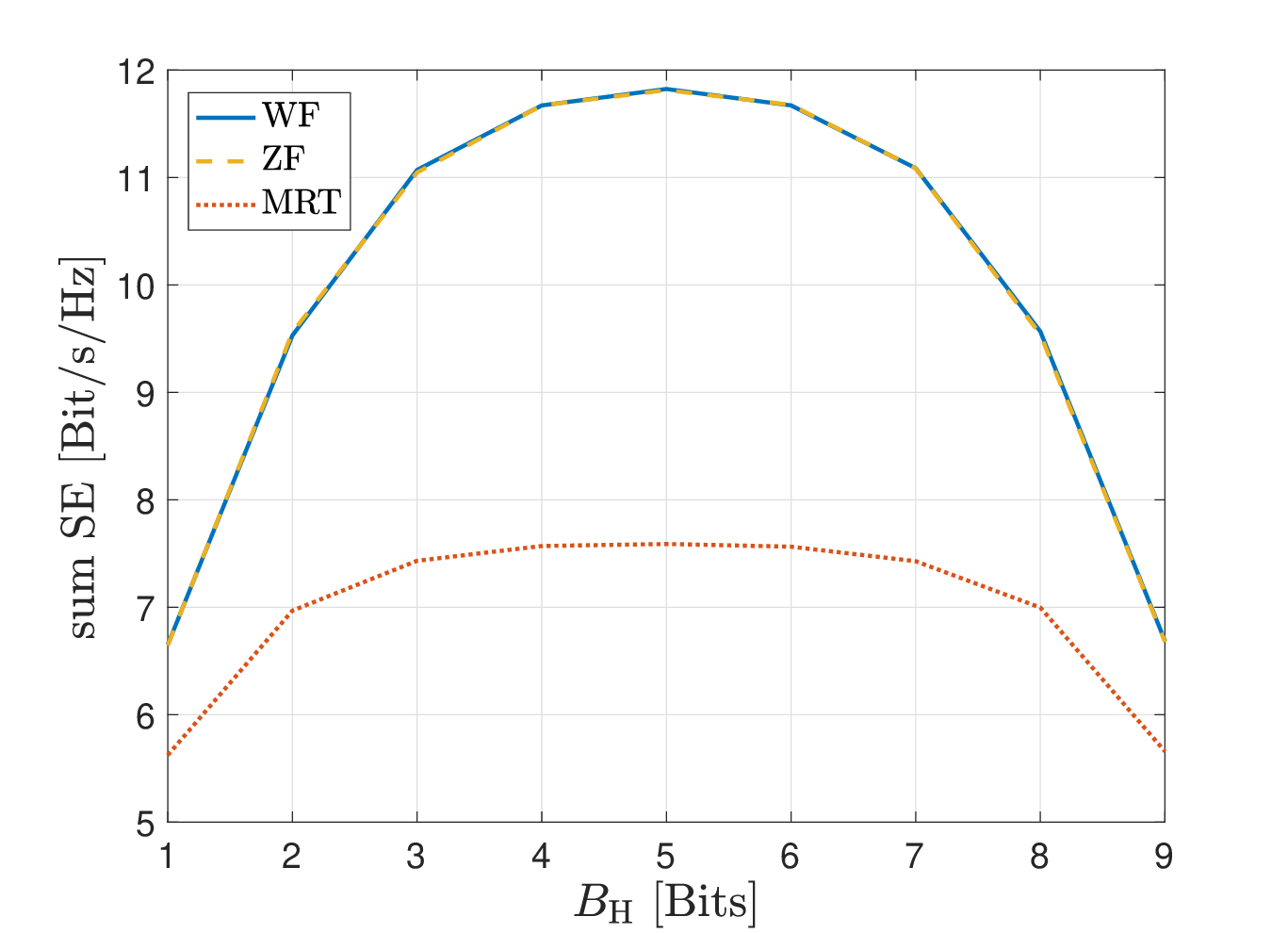}
    \caption{Sum SE versus $B_{\rm H}$ at $\text{SNR}=10$ dB.}
    \label{fig:highsnr}
\end{figure}

Fig.~\ref{fig:highsnr} evaluates the sum SE for different precoding schemes versus  $B_{\rm H}$. Here, $\text{SNR}=10$\,dB, $\bar B=10$  and ($B_{\rm P}=\bar B-B_{\rm H}$). All curves are d roughly symmetric, with the maximum at $B_{\rm H}^\star \approx 5 \quad ( B_{\rm P}\approx 5).$ Thus, at high SNR, a \emph{balanced} split between CSI and precoding bits is optimal, whereas pushing the budget to either extreme degrades performance. 
For $B_{\rm H}<5$, coarse CSI prevents effective interference suppression; for $B_{\rm H}>5$, precoding quantization dominates and residual multi-user interference grows. WF and ZF nearly coincide and attain $\approx 11.8$\,bit/s/Hz at the peak, while MRT is consistently lower ($\approx 7.6$\,bit/s/Hz) due to its inability to null inter-user interference. 
This behavior is consistent with the received-signal model in \eqref{eq:recivedDLMRT}, where at high SNR, the system becomes quantization-limited, and the effective SINR scales with the \emph{product} of CSI quality and precoding resolution—explaining why equalizing the two (i.e., $B_{\rm H}\!\approx\!B_{\rm P}$) maximizes SE under the tight budget.

\section{Conclusion}\label{sec:conclusion}
We studied joint uplink-downlink fronthaul bit allocation in a massive MU-MIMO system where the AAS estimates uplink channels, the BBU computes a downlink precoding, and both are transferred over a fronthaul link with a fixed capacity. Adopting the AQNM to describe the fronthaul quantization, we derived a unified hardening–bound SE expression. For MRT, we further obtained closed-form expressions for the SE under independent quantization noise. We verified their accuracy through Monte Carlo simulations, which explicitly show how CSI and precoding quantization affect both the coherent signal term and additive interference/noise. The closed-form MRT approximation reveals that as CSI and precoding quantization distortions decay with $B_{\rm H}$ and $B_{\rm P}$, the maximum sum SE happens by letting more bits to CSI at low SNR and becomes balanced at higher SNR. We formulated a sum–SE maximization problem subject to a fixed fronthaul budget and proposed an algorithm that finds the exact optimal bit allocation. The numerical results showed that at moderate/high SNR, there is a small CSI threshold (about $5$ bits in our setups) beyond which additional CSI resolution yields marginal gains and the performance becomes dominated by the precoding resolution. At low SNR, we should allocate approximately $80\%$ of the budget to transferring CSI, while at high SNRs, maximum sum SE is attained by balancing CSI and precoding bits (roughly $B_{\rm H}\!\approx\!B_{\rm P}$).

\bibliographystyle{IEEEtran}
\bibliography{bitallocation}
\end{document}